\documentclass[12pt]{article}
\usepackage{lscape}
\usepackage{citesort}
\usepackage{amssymb}
\usepackage{appendix}
\usepackage{multirow}

\begin{document}

\def\qq{\langle \bar q q \rangle}
\def\uu{\langle \bar u u \rangle}
\def\dd{\langle \bar d d \rangle}
\def\sp{\langle \bar s s \rangle}
\def\GG{\langle g_s^2 G^2 \rangle}
\def\Tr{\mbox{Tr}}
\def\figt#1#2#3{
        \begin{figure}
        $\left. \right.$
        \vspace*{-2cm}
        \begin{center}
        \includegraphics[width=10cm]{#1}
        \end{center}
        \vspace*{-0.2cm}
        \caption{#3}
        \label{#2}
        \end{figure}
	}
	
\def\figb#1#2#3{
        \begin{figure}
        $\left. \right.$
        \vspace*{-1cm}
        \begin{center}
        \includegraphics[width=10cm]{#1}
        \end{center}
        \vspace*{-0.2cm}
        \caption{#3}
        \label{#2}
        \end{figure}
                }

\def\ds{\displaystyle}
\def\beq{\begin{equation}}
\def\eeq{\end{equation}}
\def\bea{\begin{eqnarray}}
\def\eea{\end{eqnarray}}
\def\beeq{\begin{eqnarray}}
\def\eeeq{\end{eqnarray}}
\def\ve{\vert}
\def\vel{\left|}
\def\ver{\right|}
\def\nnb{\nonumber}
\def\ga{\left(}
\def\dr{\right)}
\def\aga{\left\{}
\def\adr{\right\}}
\def\lla{\left<}
\def\rra{\right>}
\def\rar{\rightarrow}
\def\lrar{\leftrightarrow}  
\def\nnb{\nonumber}
\def\la{\langle}
\def\ra{\rangle}
\def\ba{\begin{array}}
\def\ea{\end{array}}
\def\tr{\mbox{Tr}}
\def\ssp{{\Sigma^{*+}}}
\def\sso{{\Sigma^{*0}}}
\def\ssm{{\Sigma^{*-}}}
\def\xis0{{\Xi^{*0}}}
\def\xism{{\Xi^{*-}}}
\def\qs{\la \bar s s \ra}
\def\qu{\la \bar u u \ra}
\def\qd{\la \bar d d \ra}
\def\qq{\la \bar q q \ra}
\def\gGgG{\la g^2 G^2 \ra}
\def\gGg{m_0^2 \la \bar q q \ra}
\def\GG{\langle g_s^2 G^2 \rangle}
\def\g5{\gamma_5 \not\!q}
\def\x{\gamma_5 \not\!x}
\def\g5{\gamma_5}
\def\sb{S_Q^{cf}}
\def\sd{S_d^{be}}
\def\su{S_u^{ad}}
\def\sbp{{S}_Q^{'cf}}
\def\sdp{{S}_d^{'be}}
\def\sup{{S}_u^{'ad}}
\def\ssp{{S}_s^{'??}}

\def\sig{\sigma_{\mu \nu} \gamma_5 p^\mu q^\nu}
\def\fo{f_0(\frac{s_0}{M^2})}
\def\ffi{f_1(\frac{s_0}{M^2})}
\def\fii{f_2(\frac{s_0}{M^2})}
\def\O{{\cal O}}
\def\sl{{\Sigma^0 \Lambda}}
\def\es{\!\!\! &=& \!\!\!}
\def\ap{\!\!\! &\approx& \!\!\!}
\def\ar{&+& \!\!\!}
\def\arrr{\!\!\!\! &+& \!\!\!}
\def\ek{&-& \!\!\!}
\def\vev{&\vert& \!\!\!}
\def\kek{\!\!\!\!&-& \!\!\!}
\def\cp{&\times& \!\!\!}
\def\se{\!\!\! &\simeq& \!\!\!}
\def\eqv{&\equiv& \!\!\!}
\def\kpm{&\pm& \!\!\!}
\def\kmp{&\mp& \!\!\!}
\def\mcdot{\!\cdot\!}
\def\erar{&\rightarrow&}

% .........................................................

\def\simlt{\stackrel{<}{{}_\sim}}
\def\simgt{\stackrel{>}{{}_\sim}}

\def\ijpm{\stackrel{i=+,-}{{}_{j=+,-}}}
% .........................................................

\title{
         {\Large
                 {\bf
Radiative decays of the p--wave charmed heavy baryons 
                 }
         }
      }

\author{\vspace{1cm}\\
{\small T. M. Aliev\,$^a\!\!$ \thanks {e-mail:
taliev@metu.edu.tr}\,\,,
T. Barakat\,$^b\!\!$ \thanks {e-mail:
tbarakat@ksu.edu.sa}\,\,, M. Savc{\i}\,$^a\!\!$ \thanks
{e-mail: savci@metu.edu.tr}} \\
{\small $^a$ Physics Department, Middle East Technical University,
06531 Ankara, Turkey} \\
{\small $^b$ Physics Department, King Saud University} \\
{\small Riyadh 11451, Saudi Arabia} }
\date{}

\begin{titlepage}
\maketitle
\thispagestyle{empty}

\begin{abstract}

The radiative decays of the p--wave charmed heavy baryons to the ground state 
baryon states are studied in the framework of the light cone QCD sum rules
method. Firstly, the transition form factors that describe these
transitions are estimated, and then using these form factors the
corresponding decay widths are calculated. A comparison of our results on
the decay widths with those predicted by the other approaches existing in
literature is performed. 

\end{abstract}

%\vspace{1cm}
%~~~PACS numbers: 11.55.Hx, 13.40.Em, 14.20.Lq, 14.20.Mr \\ \\
~~~PACS numbers: 11.55.Hx, 13.40.Hq, 14.20.Lb

\end{titlepage}

\section{Introduction}

Last fifteen years is characterized (marked) by the impressive progress in hadron
physics experiments, especially on the charmed hadron physics. Many new
particles have been discovered
\cite{Rbthn01,Rbthn02,Rbthn03,Rbthn04,Rbthn05}.
Part of the new meson states could not be described by the conventional
$\qq$ picture. Initiated by these observations remarkable amount of
theoretical works  appear which lead to the conclusion that these states
could be interpreted as molecules
\cite{Rbthn06,Rbthn07,Rbthn08,Rbthn09,Rbthn10}, tetraquarks
\cite{Rbthn11,Rbthn12} or hybrids \cite{Rbthn13}. Similar situation exists
in charmed baryon sector. Some of the newly discovered charmed baryon states
\cite{Rbthn14,Rbthn15,Rbthn16} as the $J^P=\left({1\over 2}\right)^-$
 $\Lambda_c (2595)$
or the $J^P=\left({3\over 2}\right)^-$ $\Lambda_c (2625)$ can be interpreted as the
meson--baryon molecule.
At present time many negative parity baryons are discovered in the charmed
sector, while only $\Sigma_b$ baryon observed in the beauty sector. 

the study of the modes of the newly discovered negative parity heavy baryons
can be useful for establishing their nature, i.e., the conventional $qqQ$
structure, or molecular picture, or more exotic structure. The radiative
decays of these hadrons into their ground states may constitute significant
part of the total width, if the hadronic states are suppressed by the
phase--volume or if the coupling constants are small. The radiative decays
can also be useful in the determination of the quantum numbers of the
negative parity states, as well as for understanding their internal
structures.   

The present work is devoted to the study of the radiative decays of the
negative parity heavy baryons to the ground state heavy ones in framework of
the light cone QCD sum rules method (LCSR). In our analysis we have used the
background field formalism, that was introduced in \cite{Rbthn17} to
calculate the $\Sigma \to p \gamma$ decay. Note that the radiative decays
between negative (positive) parity heavy baryons is studied in the same
framework in \cite{Rbthn18} (\cite{Rbthn19}). It should also be emphasized
that the decays of the heavy hadrons are studied in framework of the LCSR in
the leading order of the heavy quark effective theory \cite{Rbthn20}. 

\section{Transition form factors of the $B_i \to B_f \gamma$ decays at
$Q^2=0$}

In the present section we will examine the transition form factors of the
$B_i \to B_f \gamma$ decays at $Q^2=0$ (for the real photon). Before
giving the details of the calculations we first present the definition of 
the transition matrix element between the initial $(B_i)$ and
final  $(B_f)$ baryons states induced by the electromagnetic current which
can be written as,
\bea
\label{ebthn01}
\la B_f(p,s) \ve j_\mu \ve B_i(p^\prime,s)\ra \es 
\bar{u}^{(s)}(p) \bigg[\bigg( \gamma_\mu - {\rlap/{q} q_\mu \over q^2} \bigg)
F_1(Q^2) - {i \sigma_{\mu\nu} q^\nu \over m_i + m_f}
F_2(Q^2) \bigg] u^{(s^\prime)}(p^\prime)\,,
\eea
where $F_1$ and $F_2$ are the Dirac and the Pauli form factors, and
$q=p_1-p_f$.
Using this matrix element for the $B_1 \to B_2 \gamma$ decay, we can easily
calculate the decay width whose expression is as,
\bea
\label{ebthn02}
\Gamma(B_{1} \to B_{2} \gamma) = {4 \alpha \ve \vec{q} \ve^3 \over
(m_{1}+ m_{2})^2 } \ve F_{2}(0) \ve^2,
\eea
where
\bea
\ve \vec{q} \ve = {m_{1}^2 - m_{2}^2 \over 2 m_{1} }\,,\nnb
\eea
is the magnitude of the photon three-momentum in the rest frame of 
the initial baryon, and 
$\alpha$ is the fine structure constant.

It follows from Eq. (\ref{ebthn02}) that
the radiative decay width is described only by the form factor $F_2(Q^2=0)$. 
The process $B_i \to B_f \gamma$ can be described by the correlation
function
\bea
\label{ebthn03}
\Pi_\mu (p,q) = - \int d^4x \int d^4y e^{i(px+qy)} \lla 0 \vel \mbox{\rm T} \{
\eta_1 (x) J_\mu^{el} (y) \bar{\eta}_2 (0) \}\ver 0 \rra \,,
\eea
where $\eta_1$ and $\eta_2$ are the interpolating currents of the
initial and final heavy baryons, $j_\mu^{el} = e_q \bar{q} \gamma_\mu q +
e_Q \bar{Q} \gamma_\mu Q$ is the electromagnetic current,
$e_q$ and $e_Q$
are the electric charges for the light and heavy quarks, respectively. 

According to the SU(3) classification the heavy baryons belong to the
symmetric sextet and antitriplet representations with respect to the light
quark exchanges. Interpolating currents corresponding to these
representations are constructed in \cite{Rbthn20} which are given as,
\bea
\eta_Q^{(S)} \es -{1\over \sqrt{2}} \epsilon^{abc} \Big\{ (q_1^{aT} C
Q^b) \gamma_5 q_2^c - (Q^{aT} C q_2^b) \gamma_5 q_1^c + \beta
\Big[(q_1^{aT} C \gamma_5 Q^b) q_2^c - (Q^{aT} C \gamma_5 q_2^b) q_1^c
\Big] \Big\}\,, \nnb \\
\eta_Q^{(A)} \es {1\over \sqrt{6}} \epsilon^{abc} \Big\{ 2 (q_1^{aT} C
q_2^b) \gamma_5 Q^c + (q_1^{aT} C Q^b) \gamma_5 q_2^c +
(Q^{aT} C q_2^b) \gamma_5 q_1^c \nnb \\
\ar \beta \Big[ 2 C(q_1^{aT} C \gamma_5 q_2^b) Q^c +
(q_1^{aT} C \gamma_5 Q^b) q_2^c +
(Q^{aT} C \gamma_5 q_2^b) q_1^c \Big] \Big\}\,, \nnb
\eea
where $a,b,c$ are the color indices, and 
$\beta$ is the arbitrary auxiliary parameter.
The light quark contents of the heavy baryons in $6$ and $\bar{3}$
representations are given in Table 1.

% .........................................................

\begin{table}[h]

\renewcommand{\arraystretch}{1.3}
\addtolength{\arraycolsep}{-0.5pt}
\small
$$
\begin{array}{|c|c|c|c|c|c|c|c|c|c|c|}
\hline \hline   
 & \Sigma_{c(b)}^{(++)+} & \Sigma_{c(b)}^{+(0)} & \Sigma_{c(b)}^{0(-)} &
             \Xi_{c(b)}^{\prime 0(-)}  & \Xi_{c(b)}^{\prime +(0)}  &
 \Omega_{c(b)}^{0(-)} & \Lambda_{c(b)}^{+(0)} & 
\Xi_{c(b)}^{0(-)}  & \Xi_{c(b)}^{+(0)} \\   
\hline \hline
q_1   & u & u & d & d & u & s & u & d & u \\
q_2   & u & d & d & s & s & s & d & s & s \\
\hline \hline 
\end{array}
$$
\caption{Light quark contents of the heavy baryons belonging to the
sextet and antitriplet representations.}
\renewcommand{\arraystretch}{1}
\addtolength{\arraycolsep}{-1.0pt}
\end{table}      

% .........................................................

The decay constant of the baryon interpolating current is defined by,
\bea
\label{ebthn04}
\la 0 \ve \eta_i \ve B(p,s)\ra \es  \lambda_i u (p)\,.
\eea
Introducing an electromagnetic background field of the plane wave 
$F_{\mu\nu} = i\left( \varepsilon_\nu^{(\lambda)} q_\mu-
\varepsilon_\mu^{(\lambda)} q_\nu \right) e^{iqx}$, the correlation
function given in Eq. (\ref{ebthn03})can be rewritten as,
\bea
\label{ebthn05}
\Pi_\mu (p,q) \varepsilon_\mu^{(\lambda)} = 
i \int d^4 xe^{i qx} \lla 0 \vel \mbox{\rm T} \{
\eta_1 (x) \bar{\eta}_2(0) \}\ver 0 \rra_F \,,
\eea
where the subscript $F$ means that all calculations are performed in the
background field $F_{\mu\nu}$. The background field method was introduced in
\cite{Rbthn21} in order to calculate the decay width of the of the $\Sigma
\to p \gamma$ process.

Let us now consider the transitions involving heavy hadrons. Along the lines
as is suggested by the usual sum rules method, we insert a complete set of
heavy baryons between the interpolating currents $\eta_1$ and $\eta_2$, and
the electromagnetic current in the correlation function given in Eq.
(\ref{ebthn03}). In doing so, there appears a problem which is absent in the
case of mesons, namely the chosen form of the interpolating current couples
not only with the positive parity ground state baryons
$J^P=\left({1\over 2}\right)^+$,
but also with a heavier baryon resonance with the negative parity 
$J^P=\left({1\over 2}\right)^-$. Many of the heavy baryons, especially the
ones involving $c$--quark have already been observed in the experiments.
Hereafter, the negative parity baryons and their masses will be denoted 
as $B^\ast$ and $m_{B^\ast}$, respectively.

The mass difference between the negative parity and ground state positive
parity baryons is about $300~MeV$, and similar mass difference is expected
for the baryons with $b$--quark. Consequently, we should take into account
the contribution of the negative parity baryons in the correlation function.
Keeping these remarks in mind, from the hadronic
side, the correlation function containing double poles can be written as,
\bea
\label{ebthn06}
\Pi_\mu(p,q) \varepsilon^\mu \es \varepsilon^\mu \sum_{\alpha,\beta}
{\la 0 \ve \eta_{Q_2} \ve B_{2\alpha} (p,s) \ra \over p^2-m_{2\alpha}^2}
\lla  B_{2\alpha} (p,s)  \ve j_\mu^{el}\ve B_{1\beta} \rra 
{\lla  B_{1\beta} (p+q,s)  \ve \bar{\eta}_{Q_1} \ve 0 \rra \over (p+q)^2 -
m_{1\beta}^2} + \cdots
\eea
where summation over $\alpha$ and $\beta$ is performed over the positive and
the negative parity baryons, and dots indicate the contributions of higher
states and continuum. 
The matrix elements in Eq. (\ref{ebthn06}) are defined as,

\bea
\label{ebthn07}
\la 0 \ve \eta_Q \ve B_2(p,s)\ra \es  \lambda_2 u (p,s)\,,\nnb \\
\la 0 \ve \eta_Q \ve B_2^\ast(p,s)\ra \es  \lambda_2^\ast \gamma_5 u^\ast (p,s)\,,\\ 
\label{ebthn08}
\la B_2(p,s) \ve j_\mu^{el} \ve B_1(p+q,s)\ra \es
\bar{u}_2 \bigg[\bigg( \gamma_\mu - {\rlap/{q} q_\mu \over q^2} \bigg)
F_1 - {i \sigma_{\mu\nu} q^\nu \over m_1 + m_2}
F_2 \bigg] u^j(p+q,s)\,,
\eea
where $\lambda_2$ $\lambda_2^\ast$ are the residues of the
baryons $B_2$ and $B_2^\ast$, and $F_1$ and $F_2$ are 
the form factors for the corresponding transitions, respectively. 

The matrix elements $\la B_2^\ast \ve j_\mu^{el} \ve B_1\ra$,    
$\la B_2 \ve j_\mu^{el} \ve B_1^\ast \ra$, and
$\la B_2^\ast \ve j_\mu^{el} \ve B_1^\ast\ra $ can easily be obtained from
Eq. (\ref{ebthn08}) with help of the replacements $\bar{u}_2 \to
\bar{u}_2^\ast \gamma_5$, $m_2 \to m_2^\ast$, $u_1 \to
\gamma_5 u_1$, and $\bar{u}_2 \to\bar{u}_2^\ast \gamma_5$,
$u_1 \to \gamma_5 u_1^\ast$, $m_1 \to m_1^\ast$,
$m_2 \to m_2^\ast$, respectively. As we already noted that the conservation of the electromagnetic
current leads to the result that the radiative decay is described only by
the form factor $F_2$.

Using the relation
\bea
\sum_s u(p)^{(s)} \bar{u}(p)^{(s)} = \rlap/{p} + m\,, \nnb 
\eea
for the summation of the Dirac spinors over the spin, and using the
transversality condition $q\mcdot \varepsilon = 0$ for the photon field, one
can easily show that only the $(p\mcdot \varepsilon)$ term is
proportional to $F_2$.

Separating the terms proportional to $F _{2-+}$, we get the following 
result for the correlation function from the phenomenological part,
\bea
\label{ebthn09}
\Pi_\mu \varepsilon^\mu = \ek
    A_{++} (\rlap/{p}+m_{2^+}) (\rlap/{p}+\rlap/{q}+m_{1^+}) (p\cdot \varepsilon) \nnb \\ 
\ar A_{--} (\rlap/{p}-m_{2^-}) (\rlap/{p}-\rlap/{q}-m_{1^-}) (p\cdot \varepsilon)\nnb \\ 
\ar A_{+-} (\rlap/{p}-m_{2^-}) (\rlap/{p}+\rlap/{q}+m_{1^+}) (p\cdot \varepsilon)\nnb \\  
\ek A_{-+} (\rlap/{p}+m_{2^+}) (\rlap/{p}+\rlap/{q}-m_{1^-}) (p\cdot \varepsilon)\nnb \\  
\eea
where
\bea
\label{ebthn10}
A_{ij} = {2 \lambda_{1^i} \lambda_{2^j} \over 
(p^2-m_{2^j}^2) [(p+q)^2
-  m_{1^i}^2]} {F_{2ij} \over (m_{1^i}+m_{2^j})}\,.
\eea

As has already been noted, in this work we study the radiative decays of the
transitions of the negative parity heavy baryons to positive heavy baryons.
The invariant function $A_{-+}$ in Eq. (\ref{ebthn09}) describes these
radiative decays under consideration. Therefore the other invariant
functions should be eliminated. It follows from Eq. (\ref{ebthn09}) that we
have four independent invariant functions, and hence we need four equations
in order to solve for the invariant function $A_{-+}$. The four equations
can be obtained from the coefficients of the structures 
$(p\mcdot \varepsilon)\rlap/{p}\rlap/{q}$,
$(p\mcdot \varepsilon)\rlap/{p}$, $(p\mcdot \varepsilon)\rlap/{q}$, and 
$(p\mcdot \varepsilon) I$.

In order to obtain the sum rules for the form factors $F_{2-+}{(0)}$ the
expression of the correlation function from the QCD side is needed. This
correlation function can be obtained can be calculated using the operator
product expansion (OPE). The OPE is performed in the deep Eucledian region
for the variables $p^2,~(p+q)^2 \!\!\ll \!m_Q^2$. In LCSR method the OPE is
performed over the twists of the nonlocal operators. The expansion of the
nonlocal operators up to twist--4 is performed in \cite{Rbthn17}, where
the four--particle contributions are neglected. Following this work we also
neglect them. The matrix elements of the nonlocal operators between the
vacuum and the photon state are parametrized in terms of the photon
distribution amplitudes (DAs). The photon DAs are studied comprehensively in
\cite{Rbthn22}, and therefore we do not present them in this work.

Equating the coefficients of the structures $(p\mcdot
\varepsilon) \rlap/{p} \rlap/{q}$,
$(p\mcdot \varepsilon) \rlap/{p}$, $(p\mcdot \varepsilon) \rlap/{q}$,
$(p\mcdot \varepsilon) I$ from the OPE side to the double hadronic dispersion
relations, we obtain the following four linearly independent equations
from which we cal solve for the form factor $F_{2-+}$,
\bea
\label{ebthn11}
-A_{++} + A_{--} + A_{+-}  -A_{-+} \es \Pi_1\,,\nnb \\
- (m_{1^+}+m_{2^+}) A_{++} - (m_{1^-}+m_{2^-})
A_{--} \!\!\! &\phantom{=}& \!\!\!\nnb \\
+ (m_{1^+}-m_{2^-}) A_{+-}
- (m_{1^-}-m_{2^+}) A_{-+} \es \Pi_2\,,\nnb \\
- m_{2^+} A_{++}  - m_{2^-} A_{--} 
-m_{2^-} A_{+-}-  m_{2^+} A_{-+}\es \Pi_3\,,\nnb \\
- m_{2^+} (m_{1^+} + m_{2^+}) A_{++} + m_{2^-}
(m_{1^-} + m_{2^-}) A_{--} \!\!\! &\phantom{=}& \!\!\!\nnb \\
- m_{2^-} (m_{1^+} - m_{2^-}) A_{+-}
+ m_{2^+} (m_{1^-} - m_{2^+}) A_{-+} \es \Pi_4\,.
\eea 

Solving the set of four equations in Eq. (\ref{ebthn11}), we can easily find
$A_{-+}$. Performing the double Borel transformation over the variables
$(p+q)^2 \to M_1^2$, $p^2 \to M_2^2$, we obtain the LCSR for the form factor
$F_{2-+}$ at the point $Q^2=0$ whose expression is given as,
\bea
\label{ebthn12}
F_{2-+}(Q^2=0) \es {m_{1^-} - m_{2^+}\over 2
\lambda_{1^-} \lambda_{2^+}}
e^{-m_{1^-}^2/M_1^2 - m_{2^-}^2/M_2^2}
\Bigg\{  {1\over (m_{1^-} + m_{1^+})
(m_{2^-} + m_{2^+})} \nnb \\
\cp\Big[m_{2^-} (m_{1^+} + m_{2^+}) m_{2^+} \Pi_1
- m_{2^-} \Pi_2
+(m_{1^+} + m_{2^+}) \Pi_3 -
\Pi_4\Big] \Bigg\} \nnb \\
\ar \int ds_1 ds_2 \rho^h (s_1,s_2) e^{-s_1/M_1^2 -s_2/M_2^2}\,,
\eea
where $\Pi_i^B$ denote the Borel--transformed invariant functions in the
coefficients of the aforementioned Lorentz structures.
The expressions of
the invariant functions $\Pi_i^B$ are quite lengthy and therefore we do not
present them in this work. The function $\rho^h (s_1,s_2)$ in the last term
is the hadronic spectral density of all excited and
continuum states with the quantum numbers of $B^\ast$.  
The hadronic spectral density is estimated
by using the quark--hadron duality ansatz. The residues of the corresponding
positive (negative) parity baryons are calculated in \cite{Rbthn23}, \cite{Rbthn24}
(\cite{Rbthn25}). In our calculations for residues we have used the results of these works.

For the processes under consideration we choose $M_1^2=M_2^2=2 M^2$
due to the fact that the masses of the initial and final baryons are quite
close to each other. The continuum subtraction can be carried out with the
help of the transformation,
\bea
(M^2)^n e^{-m_Q^2/M^2} \to {1\over \Gamma(n)} \int_{m_Q^2}^{s_0} ds
(s-m^2)^{n-1} \, e^{-s/M^2} \,,~~(n\ge 1)\,.\nnb
\eea
For the terms that are proportional to the zeroth or negative power of $M^2$ or
negative powers of $M^2$ the continuum subtraction is not performed because
these contributions are negligibly small (see for details \cite{Rbthn26}).

\section{Numerical analysis}

In order to perform the numerical analysis for the form factor $F_{2-+}(0)$
we first introduce the input parameters which we shall use in further
calculations. The main ingredient in any of the LCSR approaches in
calculating the form factors is the set of photon DAs of the particle under
consideration (in our case photon DAs). The photon DAs are obtained in
\cite{Rbthn22}, and for completeness we present their expressions in
Appendix A. The mass of the negative and positive parity baryons are taken
from \cite{Rbthn28}. The mass of the virtual $c$ quark appearing
in the heavy quark propagator is set to its value calculated in the 
$\overline{\mbox{MS}}$ scheme which is given as: $\bar{m}_c (\bar{m}_c) = 1.28 \pm 0.03~GeV$
\cite{Rbthn25}. In our analysis for the values of the quark condensates we
use $\qq(1~GeV) = -(246_{-19}^{+28}~MeV)^3$ \cite{Rbthn29,Rbthn30},
$\sp ({1~GeV}) = 0.8 \qq({1~GeV})$, $m_0^2=(0.8\pm 0.2)\,GeV^2$
\cite{Rbthn31}. The value of the magnetic susceptibility we have used is
$\chi(1\,GeV)=-0.285\,GeV^{-2}$ \cite{Rbthn32}.

The sum rule for the form factors $F_{2-+}(0)$, in addition to the
aforementioned input parameters, contain also three auxiliary parameters.
These parameters are the Borel mass parameter $M^2$, the continuum threshold
$s_0$, and the arbitrary parameter $\beta$ appearing in the expressions of
the interpolating currents. According to the QCD sum rules methodology we
need to find the so--called working regions of these parameters, where
$F_{2-+}(0)$ be insensitive to the variation of these parameters in their
working regions. Using the standard criteria of the sum rules which requires
that both power corrections and continuum contributions should sufficiently
be suppressed. The result of the analysis of the sum rules for the negative
parity heavy baryons which has been studied in \cite{Rbthn30}, leads to the
following intervals for the Borel parameter $M^2$:
\bea
&&2.5\,GeV^2 \le M^2 \le 4.0\,GeV^2,~\mbox{for~
$\Sigma_c,~\Xi_c^\prime,~\Lambda_c,~\Xi_c$}\,, \nnb \\
&&4.5\,GeV^2 \le M^2 \le 7.0\,GeV^2,~\mbox{for~
$\Sigma_b,~\Xi_b^\prime,~\Lambda_b,~\Xi_b$}\,. \nnb
\eea
The range of the values of the continuum threshold $s_0$ is determined by
imposing the condition that the sum rules should reproduce the measured
values of the form factors to within 10--15\% accuracy, from which
we obtain $11\,GeV^2 \le s_0 \le 14.0\,GeV^2$.

Our final attempt is focused on the determination of the working region of
the arbitrary parameter $\beta$. The working domain of the parameter $\beta$
is obtained by demanding that the form factor $F_{2-+}(0)$ shows good
stability with respect to the variation of $\beta$. 
%As an example, in Figs.
%(1) and (2) we present the dependence of the form factor $F_{2-+}(0)$ for
%the ?????????? transition on $\cos\theta$ in the interval $-1 \le
%\cos\theta \le 1$,  where $\beta=\tan\theta$, for
%several fixed values of the Borel parameter $M^2$ and the continuum threshold
%$s_0$. It follows from these figures that the behavior of the form factor
%$F_{2-+}(0)$ seems to be practically insensitive to the variation in
%$\cos\theta$ when it varies in the interval $???\le \cos\theta \le ????$.
We have gone through analysis for all possible transitions, and the form factors
$ F _{2-+}(0) $ seems to be practically insensitive to the variation in 
$ \cos \theta $ in the domain $ -1 \leq \cos \theta \leq -0.5 $ and
observed that this working region of $\cos\theta$ is practically common for
all transitions. Our final results for the form factor $F_{2-+}(0)$ for all
transition channels are given below,
\bea
% ---------------------------------------------
% SEXTET to SEXTET
F_{2-+}(0) = \left\{ \begin{array}{ll}
         %  & \mbox{(Sextet $\to$ Sextet)}  \\
(0.5\pm 0.05) & \Sigma_c^{0\ast}(2792)     \to \Sigma_c^{0}     \gamma \\
(-1.0\pm 0.2) & \Sigma_c^{+\ast}(2792)     \to \Sigma_c^{+}     \gamma \\
(-3.5\pm 0.05) & \Sigma_c^{++\ast}(2792)    \to \Sigma_c^{++}    \gamma \\
(2.5\pm 0.5) & \Xi_c^{\prime 0\ast}(2870) \to \Xi_c^{\prime 0} \gamma \\
(-1\pm 0.1) & \Xi_c^{\prime +\ast}(2870) \to \Xi_c^{\prime +} \gamma \\
% ---------------------------------------------
% SEXTET to TRIPLET
         %  & \mbox{(Sextet $\to$ Triplet)}  \\
(2.5\pm 0.5) & \Sigma_c^{+\ast}(2792)     \to \Lambda_c^{+}    \gamma \\
(-0.3\pm 0.05) & \Xi_c^{\prime 0\ast}(2870) \to \Xi_c^{0}        \gamma \\
(2\pm 0.3) & \Xi_c^{\prime +\ast}(2870) \to \Xi_c^{+}        \gamma \\
% ---------------------------------------------
% TRIPLET to TRIPLET
          % & \mbox{(Triplet $\to$ Triplet)}  \\
(0.3\pm 0.07) & \Xi_c^{0\ast}(2790)         \to \Xi_c^{0}       \gamma \\
(3\pm 0.5) & \Xi_c^{+\ast}(2790)         \to \Xi_c^{+}       \gamma \\
(2.5\pm 0.4) & \Lambda_c^{+\ast}(2592)     \to \Lambda_c^{+}   \gamma \\
% ---------------------------------------------
% TRIPLET to SEXTET
           %& \mbox{(Triplet $\to$ Sextet)}  \\
(3\pm 0.5) & \Lambda_c^{+\ast}(2592)     \to \Sigma_c^{+}     \gamma \\ 
(0.25\pm 0.05) & \Xi_c^{0\ast}(2790)         \to \Xi_c^{\prime 0} \gamma \\ 
(2.5\pm 0.5) & \Xi_c^{+\ast}(2790)         \to \Xi_c^{\prime +} \gamma \\ 
\end{array} \right. \nnb
\eea

The errors in the results are estimated by changing various input parameters
within their working regions, and by also taking into consideration the
resulting separate errors of the form factors $ F _{2-+}(0) $ quadratically.

Having obtained the values of the form factors $F_{2-+}(0)$, we can
calculate the decay widths of the decays under consideration. The values of
the decay widths are presented in Table 2. For completeness we also present
the values of the decay widths for the same transitions obtained in
framework of the other approaches such as the coupled channel dynamically
generated model \cite{Rbthn33}, relativistic quark model \cite{Rbthn34},
light cone sum rules method \cite{Rbthn35}, chiral perturbation theory
\cite{Rbthn36}, bound state picture \cite{Rbthn37}, and constituent quark
model \cite{Rbthn38}.

% .........................................................

%\landscape

\newcommand{\rb}[1]{\raisebox{1.5ex}[0pt]{#1}}

\begin{table}[thb]
\centering
\renewcommand{\arraystretch}{1.3}
\addtolength{\arraycolsep}{-0.5pt}

\small
$$
\begin{array}{|l|c|c|c|c|c|c||c|c|}
\hline \hline
%  \multirow{2}{*}{ }        &\multicolumn{2}{c||}{\cite{Rozd16}}   
%                            &\multicolumn{4}{c|}{\mbox{present work}}
  
%\cline{1-7} 
&
\multirow{2}{*}{\mbox{This work}} &                     
\multirow{2}{*}{\cite{Rbthn33}}   &                     
\multirow{2}{*}{\cite{Rbthn34}}   & 
\multirow{2}{*}{\cite{Rbthn35}}   &  
\multirow{2}{*}{\cite{Rbthn36}}   & 
\multirow{2}{*}{\cite{Rbthn37}}   & 
\multicolumn{2}{c|}{\cite{Rbthn38}}      
\\ \cline{8-9}
                                   &
             \multicolumn{1}{c|}{} &
             \multicolumn{1}{c|}{} &
             \multicolumn{1}{c|}{} &
             \multicolumn{1}{c|}{} &
             \multicolumn{1}{c|}{} &
             \multicolumn{1}{c||}{}&
             \multicolumn{1}{c|}{\mbox{Case A}} &
             \multicolumn{1}{c|}{\mbox{Case B}}
\\ \hline
% SEXTET to SEXTET
%\mbox{(Sextet $\to$ Sextet)}& & & & & & & & \\
\Sigma_c^{0\ast}(2792)     \to \Sigma_c^{0}      \gamma  & 9(1 \pm 0.2) & 9 & - & - & - & - & 5.02 & - \\
\Sigma_c^{+\ast}(2792)      \to \Sigma_c^{+}      \gamma  & 36(1\pm 0.4)  & 29 & - & - & - & - & 0.92 & - \\
\Sigma_c^{++\ast}(2792)    \to \Sigma_c^{++}     \gamma  & 440(1\pm 0.25) &  51  & - & - & - & & 8.54 & - \\
\Xi_c^{\prime 0\ast}(2870) \to \Xi_c^{\prime 0}  \gamma  & 132(1\pm 0.2)  &    & & & & & & \\
\Xi_c^{\prime +\ast}(2870) \to \Xi_c^{\prime +}  \gamma  & 21 (1\pm 0.2)  &    & & & & & & \\ \hline
% SEXTET to TRIPLET
%\mbox{(Sextet $\to$ Triplet)}                          &   & & & & & & & \\
\Sigma_c^{+\ast}(2792)     \to \Lambda_c^{+}     \gamma  & 701(1\pm 0.4) & 35 & - & & - & - & 52.1 & - \\
\Xi_c^{\prime 0\ast}(2870) \to \Xi_c^{0}         \gamma  & 4.8(1\pm 0.3) & & & & & & & \\
\Xi_c^{\prime +\ast}(2870) \to \Xi_c^{+}         \gamma  & 214 (1\pm 0.3) & & & & & & & \\ \hline
%TRIPLET to TRIPLET
%\mbox{(Triplet $\to$ Triplet)}                         & & & & & & & & \\
\Xi_c^{0\ast}(2790)        \to \Xi_c^{0}         \gamma  & 2.7(1\pm 0.3) & 117 & - & - & - & - & 263 & 5.57 \\
\Xi_c^{+\ast}(2790)        \to \Xi_c^{+}         \gamma  & 265(1\pm 0.4) & 246 & - & - & - & - & 4.65 & 1.39 \\
\Lambda_c^{+\ast}(2592)    \to \Lambda_c^{+}     \gamma  & 189(1\pm 0.3)  & 278 & 115 \pm 1  & 36 & 0  &  16 & 0.26 & 1.59 \\ \hline
% TRIPLET to SEXTET
%\mbox{(Triplet $\to$ Sextet)}                          & & & & & & & & \\
\Lambda_c^{+\ast}(2592)    \to \Sigma_c^{+}      \gamma  &  29(1\pm 0.3) & 2   & 77 \pm 1   & 11 & 71 &   - & 0.45 & 41.6 \\
\Xi_c^{0\ast}(2790)   \to \Xi_c^{\prime 0}  \gamma  & 0.54(1\pm 0.4) & 1 & - & - & - & - & 0 & 0 \\
\Xi_c^{+\ast}(2790)   \to \Xi_c^{\prime +}  \gamma  & 54(1\pm 0.4) & 1 & - & - & - & - & 1.43 & 128\\
\hline \hline
\end{array}
$$

\caption{The values of radiative decay widths (in units of $keV$).}

\renewcommand{\arraystretch}{1}
\addtolength{\arraycolsep}{-1.0pt}

\end{table}

%\endlandscape

% .........................................................

Reviewing the values of the decay widths given in Table 1, we see that between 
our results and predictions coming from other approaches on many channels, 
there are considerable differences. Therefore, measurement of decay widths 
can serve as a good tool for ``choosing the right model". 
From Table 1, we conclude that, our approach predict quite large values, 
especially for the decay channels and these channels 
($ \Sigma_c^{++\ast}    \to \Sigma_c^{++}\gamma $, $ \Xi_c^{\prime 0\ast} \to 
\Xi_c^{\prime 0}\gamma $, $ \Sigma_c^{+\ast}     \to \Lambda_c^{+}\gamma $, 
$ \Xi_c^{\prime +\ast} \to \Xi_c^{+} \gamma$, $ \Xi_c^{+\ast} \to 
\Xi_c^{+}\gamma  $, and $ \Lambda_c^{+\ast}    \to \Lambda_c^{+} \gamma$) 
can be observed in the new future in experiments conducting at accelerators.
Our final remark on this section is that, the decay widths for corresponding 
beauty baryons can be obtained from these calculations by performing relevant 
replacements

\section{Conclusion}

In this work we study the radiative decays of the negative parity heavy baryons to
the ground state positive parity heavy baryons in framework of the LCSR
method. We obtain that the transitions $ \Sigma_c^{++\ast}    \to 
\Sigma_c^{++}\gamma $, $ \Xi_c^{\prime 0\ast} \to \Xi_c^{\prime 0}\gamma $,
$ \Sigma_c^{+\ast}     \to \Lambda_c^{+}\gamma $, $ \Xi_c^{\prime +\ast} \to 
\Xi_c^{+} \gamma$, $ \Xi_c^{+\ast}        \to \Xi_c^{+}\gamma  $, and 
$ \Lambda_c^{+\ast}    \to \Lambda_c^{+} \gamma$
have sizable decay width values. Our results show that the decay widths
calculated in this work are quite different from the ones predicted by the
other approaches. The apparently sizable values of the decay widths we
obtain indicate that these radiative decays can serve a very useful tool for
gaining information about the properties of the negative parity heavy
baryons.         

\section*{Acknowledgment}
	One of the us, T. Barakat, thanks to the
International Scientific Partnership Program ISPP at the King Saud
University for funding his research work through ISPP No: 0038.  

\clearpage

\section*{Appendix A: Photon distribution amplitudes (DA's)}
In the present Appendix, for completeness we present the definitions of the 
photon DA's obtained in \cite{Rbthn22}.
\bea
&&\langle \gamma(q) \vert  \bar q(x) \sigma_{\mu \nu} q(0) \vert  0
\rangle  = -i e_q \qq (\varepsilon_\mu q_\nu - \varepsilon_\nu
q_\mu) \int_0^1 du e^{i \bar u qx} \left(\chi \varphi_\gamma(u) +
\frac{x^2}{16} \mathbb{A}  (u) \right) \nnb \\ &&
-\frac{i}{2(qx)}  e_q \qq \left[x_\nu \left(\varepsilon_\mu - q_\mu
\frac{\varepsilon x}{qx}\right) - x_\mu \left(\varepsilon_\nu -
q_\nu \frac{\varepsilon x}{q x}\right) \right] \int_0^1 du e^{i \bar
u q x} h_\gamma(u)
\nnb \\
&&\langle \gamma(q) \vert  \bar q(x) \gamma_\mu q(0) \vert 0 \rangle
= e_q f_{3 \gamma} \left(\varepsilon_\mu - q_\mu \frac{\varepsilon
x}{q x} \right) \int_0^1 du e^{i \bar u q x} \psi^v(u)
\nnb \\
&&\langle \gamma(q) \vert \bar q(x) \gamma_\mu \gamma_5 q(0) \vert 0
\rangle  = - \frac{1}{4} e_q f_{3 \gamma} \epsilon_{\mu \nu \alpha
\beta } \varepsilon^\nu q^\alpha x^\beta \int_0^1 du e^{i \bar u q
x} \psi^a(u)
\nnb \\
&&\langle \gamma(q) | \bar q(x) g_s G_{\mu \nu} (v x) q(0) \vert 0
\rangle = -i e_q \qq \left(\varepsilon_\mu q_\nu - \varepsilon_\nu
q_\mu \right) \int {\cal D}\alpha_i e^{i (\alpha_{\bar q} + v
\alpha_g) q x} {\cal S}(\alpha_i)
\nnb \\
&&\langle \gamma(q) | \bar q(x) g_s \widetilde G_{\mu \nu} i \gamma_5 (v
x) q(0) \vert 0 \rangle = -i e_q \qq \left(\varepsilon_\mu q_\nu -
\varepsilon_\nu q_\mu \right) \int {\cal D}\alpha_i e^{i
(\alpha_{\bar q} + v \alpha_g) q x} \widetilde {\cal S}(\alpha_i)
\nnb \\
&&\langle \gamma(q) \vert \bar q(x) g_s \widetilde G_{\mu \nu}(v x)
\gamma_\alpha \gamma_5 q(0) \vert 0 \rangle = e_q f_{3 \gamma}
q_\alpha (\varepsilon_\mu q_\nu - \varepsilon_\nu q_\mu) \int {\cal
D}\alpha_i e^{i (\alpha_{\bar q} + v \alpha_g) q x} {\cal
A}(\alpha_i)
\nnb \\
&&\langle \gamma(q) \vert \bar q(x) g_s G_{\mu \nu}(v x) i
\gamma_\alpha q(0) \vert 0 \rangle = e_q f_{3 \gamma} q_\alpha
(\varepsilon_\mu q_\nu - \varepsilon_\nu q_\mu) \int {\cal
D}\alpha_i e^{i (\alpha_{\bar q} + v \alpha_g) q x} {\cal
V}(\alpha_i) \nnb \\ && \langle \gamma(q) \vert \bar q(x)
\sigma_{\alpha \beta} g_s G_{\mu \nu}(v x) q(0) \vert 0 \rangle  =
e_q \qq \left\{
        \left[\left(\varepsilon_\mu - q_\mu \frac{\varepsilon x}{q x}\right)\left(g_{\alpha \nu} -
        \frac{1}{qx} (q_\alpha x_\nu + q_\nu x_\alpha)\right) \right. \right. q_\beta
\nnb \\ && -
         \left(\varepsilon_\mu - q_\mu \frac{\varepsilon x}{q x}\right)\left(g_{\beta \nu} -
        \frac{1}{qx} (q_\beta x_\nu + q_\nu x_\beta)\right) q_\alpha
\nnb \\ && -
         \left(\varepsilon_\nu - q_\nu \frac{\varepsilon x}{q x}\right)\left(g_{\alpha \mu} -
        \frac{1}{qx} (q_\alpha x_\mu + q_\mu x_\alpha)\right) q_\beta
\nnb \\ &&+
         \left. \left(\varepsilon_\nu - q_\nu \frac{\varepsilon x}{q.x}\right)\left( g_{\beta \mu} -
        \frac{1}{qx} (q_\beta x_\mu + q_\mu x_\beta)\right) q_\alpha \right]
   \int {\cal D}\alpha_i e^{i (\alpha_{\bar q} + v \alpha_g) qx} {\cal T}_1(\alpha_i)
\nnb \\ &&+
        \left[\left(\varepsilon_\alpha - q_\alpha \frac{\varepsilon x}{qx}\right)
        \left(g_{\mu \beta} - \frac{1}{qx}(q_\mu x_\beta + q_\beta x_\mu)\right) \right. q_\nu
\nnb \\ &&-
         \left(\varepsilon_\alpha - q_\alpha \frac{\varepsilon x}{qx}\right)
        \left(g_{\nu \beta} - \frac{1}{qx}(q_\nu x_\beta + q_\beta x_\nu)\right)  q_\mu
\nnb \\ && -
         \left(\varepsilon_\beta - q_\beta \frac{\varepsilon x}{qx}\right)
        \left(g_{\mu \alpha} - \frac{1}{qx}(q_\mu x_\alpha + q_\alpha x_\mu)\right) q_\nu
\nnb \\ &&+
         \left. \left(\varepsilon_\beta - q_\beta \frac{\varepsilon x}{qx}\right)
        \left(g_{\nu \alpha} - \frac{1}{qx}(q_\nu x_\alpha + q_\alpha x_\nu) \right) q_\mu
        \right]
    \int {\cal D} \alpha_i e^{i (\alpha_{\bar q} + v \alpha_g) qx} {\cal T}_2(\alpha_i)
\nnb \\ &&+
        \frac{1}{qx} (q_\mu x_\nu - q_\nu x_\mu)
        (\varepsilon_\alpha q_\beta - \varepsilon_\beta q_\alpha)
    \int {\cal D} \alpha_i e^{i (\alpha_{\bar q} + v \alpha_g) qx} {\cal T}_3(\alpha_i)
\nnb \\ &&+
        \left. \frac{1}{qx} (q_\alpha x_\beta - q_\beta x_\alpha)
        (\varepsilon_\mu q_\nu - \varepsilon_\nu q_\mu)
    \int {\cal D} \alpha_i e^{i (\alpha_{\bar q} + v \alpha_g) qx} {\cal T}_4(\alpha_i)
                        \right\} \nnb \\
&&\langle \gamma(q) \vert \bar q(x) e_q F_{\mu\nu} (vx) q(0) \vert 0 \rangle =
-i e_q \qq \left(\varepsilon_\mu q_\nu - \varepsilon_\nu q_\mu \right)
\int {\cal D}\alpha_i e^{i (\alpha_{\bar q} + v \alpha_g) q x}
{\cal S}^\gamma (\alpha_i)\nnb \\
&&\langle \gamma(q) \vert \bar q(x) \sigma_{\alpha\beta} F_{\mu\nu}
(vx) q(0) \vert 0 \rangle = e_q \qq {1\over qx} (q_\alpha x_\beta - q_\beta x_\alpha)
\left(\varepsilon_\mu q_\nu - \varepsilon_\nu q_\mu \right) \nnb \\
&&\times \int {\cal D}\alpha_i e^{i (\alpha_{\bar q} + v \alpha_g) q x}
{\cal T}_4^\gamma (\alpha_i)\nnb \,,
\eea
where $\varphi_\gamma(u)$ is the leading twist--2, $\psi^v(u)$,
$\psi^a(u)$, ${\cal A}$ and ${\cal V}$ are the twist--3, and
$h_\gamma(u)$, $\mathbb{A}$, ${\cal S}$, $\widetilde{\cal S}$,
${\cal S}^\gamma$,
${\cal T}_i$ ($i=1,~2,~3,~4$), ${\cal T}_4^\gamma$ are the
twist--4 photon DAs, $\chi$ is the magnetic susceptibility, and
the measure ${\cal D} \alpha_i$ is defined as
\bea
\int {\cal D} \alpha_i = \int_0^1 d \alpha_{\bar q} \int_0^1 d
\alpha_q \int_0^1 d \alpha_g \delta(1-\alpha_{\bar
q}-\alpha_q-\alpha_g)~.\nnb
\eea

\clearpage

\end{document}